\documentclass[10pt,twocolumn]{article}

\usepackage{graphicx}
\usepackage{amsfonts}
\usepackage{amsmath}
\usepackage{amssymb}
\usepackage{mathrsfs} 
\usepackage{lipsum}
\usepackage{color}
\usepackage{siunitx}
\usepackage{authblk}
\usepackage[margin=2cm]{geometry}
\usepackage{mathtools, cuted}

\usepackage{bm}
\usepackage{amsmath}

\title{Arresting bubble coarsening: A two-bubble experiment to investigate grain growth in presence of surface elasticity}

\author[1]{A. Salonen}
\author[2]{C. Gay} 
\author[1]{A. Maestro}
\author[1]{W. Drenckhan}
\author[1]{E. Rio}

\affil[1] {Universit\'e Paris-Sud, Laboratoire de Physique des Solides, UMR8502, Orsay, F-91405}
\affil[2] {Universit\'e Paris~Diderot--Paris~7
Mati\`ere et Syst\`emes Complexes (CNRS UMR 7057),
B\^atiment Condorcet, Case courrier 7056, 75205 Paris Cedex 13}
\affil[3] {Present address: Biological and Soft Systems, University of Cambridge, UK.}

\date{\today}

\begin{document}

\newcommand{\hs}{\hspace{0.7cm}}
\newcommand{\be}{\begin{equation}}
\newcommand{\ee}{\end{equation}}
\newcommand{\bee}{\begin{eqnarray}}
\newcommand{\eee}{\end{eqnarray}}
\newcommand{\fin}{\nonumber\\}
\newcommand{\hide}[1]{}

\newcommand{\elasticity}{E}
\newcommand{\area}{A}
\newcommand{\pressure}{p}
\newcommand{\topen}{t_0}
\newcommand{\lcap}{\ell_{\rm cap}}


\maketitle

\abstract{Many two-phase materials suffer from grain-growth due to the energy cost 
which is associated with the interface that separates both phases. 
While our understanding of the driving forces and the dynamics 
of grain growth in different materials is well advanced by now, 
current research efforts address the question of how this process may be slowed down, 
or, ideally, arrested. 
We use a model system of two bubbles to explore 
how the presence of a finite surface elasticity may interfere 
with the coarsening process and the final grain size distribution. 
Combining experiments and modelling in the analysis of the evolution of two bubbles, 
we show that clear relationships can be predicted between the surface tension, 
the surface elasticity and the initial/final size ratio of the bubbles.
We rationalise these relationships by the introduction of a modified \emph{Gibbs criterion}.
Besides their general interest, the present results 
have direct implications for our understanding of foam stability.
}

\section{Introduction}

Materials consisting of grains separated by well-defined interfaces are ubiquitous. 
Examples include polycrystalline solids ~\cite{Novikov_BOOK_1996}, 
magnetic garnet films ~\cite{Weaire_Bolton_1991}, 
two-phase ferrofluidic mixtures ~\cite{Elias_Flament_1997},  
superconducting magnetic froths ~\cite{Prozorov_Fidler_2008}, 
foams ~\cite{Weaire_Hutzler_1999,Cantat_BOOK_2013} 
or emulsions ~\cite{Binks_BOOK_1998}. 
In such systems, the positive energy associated with the interfaces 
is the driving force of a characteristic grain growth or "coarsening" process 
by which smaller grains tend to disappear 
while larger grains grow, 
leading to a progressive reduction of the overall interfacial energy
and to characteristic asymptotic grain size distributions. 
 
While our understanding of the main mechanisms of grain growth in these different systems 
has advanced significantly, 
much effort is now dedicated to the question of how this grain growth 
may be controlled or, ideally, completely arrested. 
Since the historic work by the metallurgist S. C. Smith ~\cite{Smith_1981}, 
liquid foams have served repeatedly as model systems for related questions. 
We return here to this model system in order to tackle the question 
of how grain growth may be arrested by the presence 
of a surface elasticity ~\cite{Gibbs1993,Kloek2001}. In this case, the 
interfaces have a surface tension $\gamma$ which
depends on the interfacial  area $\area$ leading to an additional resistance to grain growth. 
This resistance is characterised by a dilational elastic surface modulus $\elasticity$ defined as
\bee
\elasticity=\frac{\partial \gamma}{\partial \ln \area}.
\label{SurfaceElasticity}
\eee
\begin{figure}[h!]
\begin{center}
\resizebox{1.0\columnwidth}{!}{%
\includegraphics{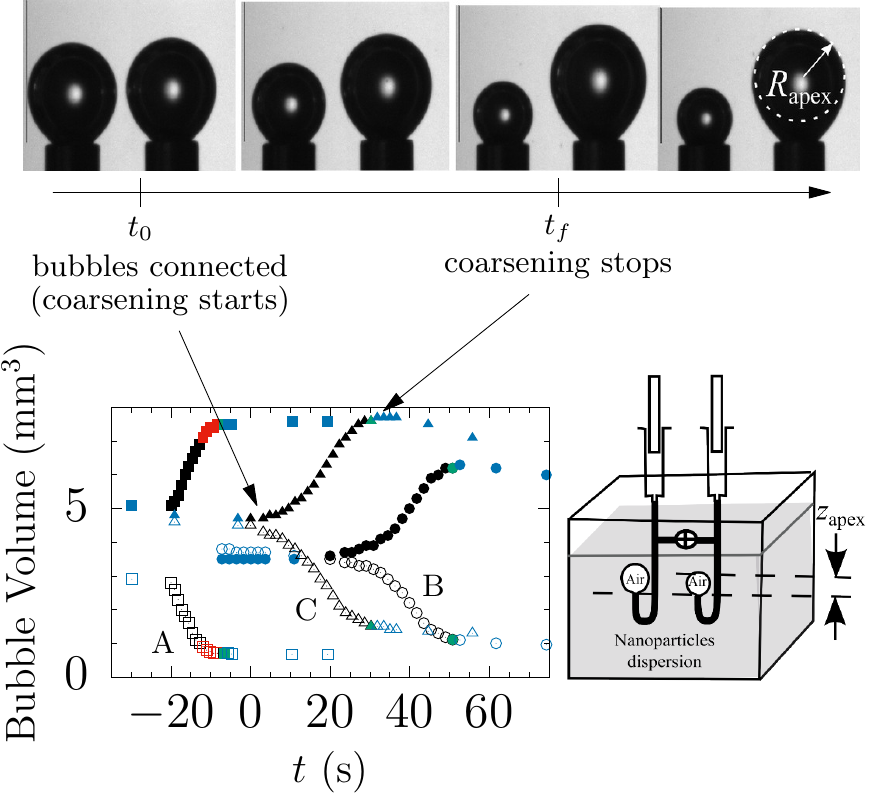}
}
\end{center}
\caption{
Experimental set-up and bubble volume evolution.
{\em Top:} Photographs of the bubbles at four different times 
(the distance between both needle axes is about 7.5~mm.
{\em Bottom right:} Setup. 
Two bubbles are prepared in a nanoparticle dispersion and left to equilibrate.
A valve connects both bubbles. It is opened at $\topen$. 
Gas exchange between both bubbles occurs 
until their volumes reach their final values.
{\em Bottom left:} Evolution of the bubble volumes (mm$^3$) for three different experiments. 
Arbitrary time shifts have been applied for clarity.
Experiment A (resp B and C) is represented by squares (resp. circles and triangles). 
Filled symbols are for the growing bubbles while open symbols stand for the shrinking bubbles.
The present study focuses on the black data points 
corresponding to monotonically changing volumes and surface areas.
The last such point is marked in green
and the corresponding time is called $t_f$.
Points outside this range are represented in blue.
Red points for experiment A correspond to a regime of larger surface elasticity,
see Fig.~\ref{fig:courbes:experimentales}b).
}
\label{fig:Setup}
\end{figure}
In this context, a classical prediction for the growth arrest
was given by Gibbs~\cite{Gibbs1993}. 
It consists in considering a single spherical bubble of radius $R$ and of surface area $A=4\pi R^2$
in a liquid at constant pressure $p_{\rm liq}$. 
In the absence of a surface elasticity, the gas pressure $p_{\rm gas}$ inside the bubble
exceeds that in the liquid, the corresponding pressure drop being given
by the Young-Laplace law:
\begin{equation}
\label{eq:laplace}
\Delta p = p_{\rm gas} - p_{\rm liq} = 2\gamma/R.
\end{equation}
This pressure drop increases with decreasing bubble size 
($\partial \Delta p/\partial R < 0$), leading to an accelerated bubble dissolution. 
Gibbs showed that this evolution can be not only slowed down but even fully arrested 
by the presence of a sufficiently high surface elasticity.
Indeed, Eqs.~(\ref{SurfaceElasticity}-\ref{eq:laplace}) imply that
$\partial \Delta p/\partial R = \frac{2}{R^2}\,(2\elasticity-\gamma)$.
Hence, when $\elasticity/\gamma \geq 1/2$, 
which is now known as the \emph{Gibbs criterion} \cite{Gibbs1993}, the bubble evolution is stopped ($\frac{\partial P}{\partial R}<0$).  
This criterion has been verified by simulations more recently \cite{Kloek2001}.
 
In the case of a single bubble, such a surface elasticity arises naturally 
when working with stabilising agents which are irreversibly adsorbed to the gas/liquid interface. 
Systems which are very much “en vogue” for this purpose consist of bubbles 
stabilised by nano- or micron-sized particles~\cite{Gonzenbach2006,Horozov2008,Dickinson2010} 
or special proteins~\cite{Blijdenstein2010}. 
In this case, during the shrinking process,
the agents are compacted at the interface,
hence the surface tension $\gamma$ decreases which may lead to $\elasticity / \gamma \geq 1/2$.

Many irreversibly adsorbed systems have been used to stabilise foams, which have indeed proven to stop coarsening, with a surprisingly good agreement with the Gibbs criterion~\cite{Cervantes2008,Stocco2009,Blijdenstein2010,Georgieva2009,Cox2009}. 
However, many questions remain as to how the behaviour of a single bubble 
can be related to that of a complex foam which contains bubbles of different sizes – 
since some of them will shrink and others will grow. 
Moreoever, the coarsening process can be slowed down or arrested for other reasons, 
the bubble surfaces might become impermeable to gas arresting the diffusion process 
or the shear viscoelasticity of the bubbles might stop rearrangements which would hinder coarsening. 
Thus, it is not clear at this stage how to provide a reliable criterion for the coarsening to stop in a foam, 
and to predict the final bubble size distribution~\cite{Kloek2001}. 
  
In order to understand the mechanisms of the arrest of coarsening it is necessary 
to carry out model experiments which allow to discriminate between the different processes. 
Very recently studies were carried out on a single bubble in a pressure-controlled solution 
\cite{Taccoen2016}. The pressure of the solution controls the concentration of the dissolved gas 
and hence its partial pressure, and is used to explore the stability to coarsening of isolated bubbles. 
The authors \cite{Taccoen2016} propose a model taking into account the energy dissipation 
due to the buckling of the interface in contradiction with the Gibbs criterion. 

To get one step closer to foams, we propose, in this article, a two bubble experiment, 
where the bubbles are connected by a tube. This allows to incorporate a first degree of polydispersity 
(the size difference between both bubbles). Moreover, in this experiment, the coarsening 
is driven by the pressure difference between the two bubbles as in foams \cite{Webster2001}, 
rather than by the liquid/bubble pressure difference as in a single bubble experiment. 
In the following we thus compare successfully a simple two bubble experiment presented 
in the next section with a model developped in the third section. 
In a last section, we show that the Gibbs criterion can be recovered theoretically in specific conditions.
 
\section{Two bubble experiments}
The set-up we are interested in is schematised in FIG.~\ref{fig:Setup}. 
A small bubble (1) and a big bubble (2) are connected by a tube. 
More precisely, two syringes are immersed in the same solution and their outlets are positioned 
at the same altitude. 
The gaz is pushed through the syringes manually to generate a bubble at the syringe outlet
(FIG.~\ref{fig:Setup}, bottom right). Both syringe outlets are connected by a tube 
which is initially closed by a valve. The bubble evolution is recorded 
with a video camera (FIG.~\ref{fig:Setup}, top). 
The pictures are then treated by the software included in the Tracker device (Teclis, France) 
to extract the volume, surface area, surface tension, apex altitude and apex radius of curvature 
of each bubble as a function of time. 
The experimental setup was tested with bubbles 
made in a solution sodium dodecyl sulphate (SDS) at 10 mM purchased from Sigma Aldrich. 
The transfer of gas from the smaller to the larger bubble occurs very rapidly 
once the tubes are connected indicating that the resistance of the tubing is small.

The bubbles are created in an aqueous dispersion of silica nanoparticles 
(Ludox, TMA from Sigma Aldrich) with a 25 nm diameter. 
The dispersions are stable as the silica particles 
are negatively charged at the pH used, which is close to 7. 
In order to make the particle surface active, 
a positively charged surfactant, cetyl trimethyl ammonium bromide (CTAB), is added \cite{Maestro2014}. 
The surfactant adsorbs onto the surface of the particles and makes them partially hydrophobic. 
The CTAB is also purchased from Sigma and used without further purification. 
The samples are prepared in Milli-Q water (conductivity 18.2 M$\Omega$.cm$^{-1}$) 
with 1 mM of NaBr to promote adsorption (in line with previous experiments \cite{Maestro2014}). 
The experiments are carried out with 1 wt\% silica particles and 10$^{-4}$~M CTAB. 
In this system the surface elasticity is constant at sufficiently large bubble shrinkage 
and should be high enough to fulfill the Gibbs criterion \cite{Maestro2014}. 
We varied the initial size of the bubbles, which impacts the final state, 
as we will show in the following. 
We discuss here in details three experiments with different initial bubble sizes.

FIG.~\ref{fig:Setup} (bottom, left) displays the evolution of the volumes $V_1$ (open symbols) 
and $V_2$ (filled symbols) of the small and big bubbles respectively 
(different symbols correspond to three different data set). 
Initially, the bubbles are not connected, yet the 
volumes $V_1$ and $V_2$ decrease weakly with time (see FIG.~\ref{fig:Setup}, bottom left). 
This indicates a slow dissolution of the bubbles into the bath. Then, 
at some time $t_0$ (note that for each experiment, the times have been shifted arbitrarily for clarity),
the valve is opened and the gas is free to flow from one bubble to the other.
The smaller bubble then shrinks
while the bigger one grows, the dynamics of exchange being set by the surface viscosity. 
This coarsening behaviour stops 
after typically $20$ to $30$~s
(at a time we call $t_f$)
and the bubbles reach a stable final volume - apart from the drift due to slow dissolution.
Note that the bubbles are still connected by a tube at this time. 
This shows that the coarsening has stopped.

\begin{figure}[h!]
\begin{center}
\resizebox{1.0\columnwidth}{!}{
\includegraphics{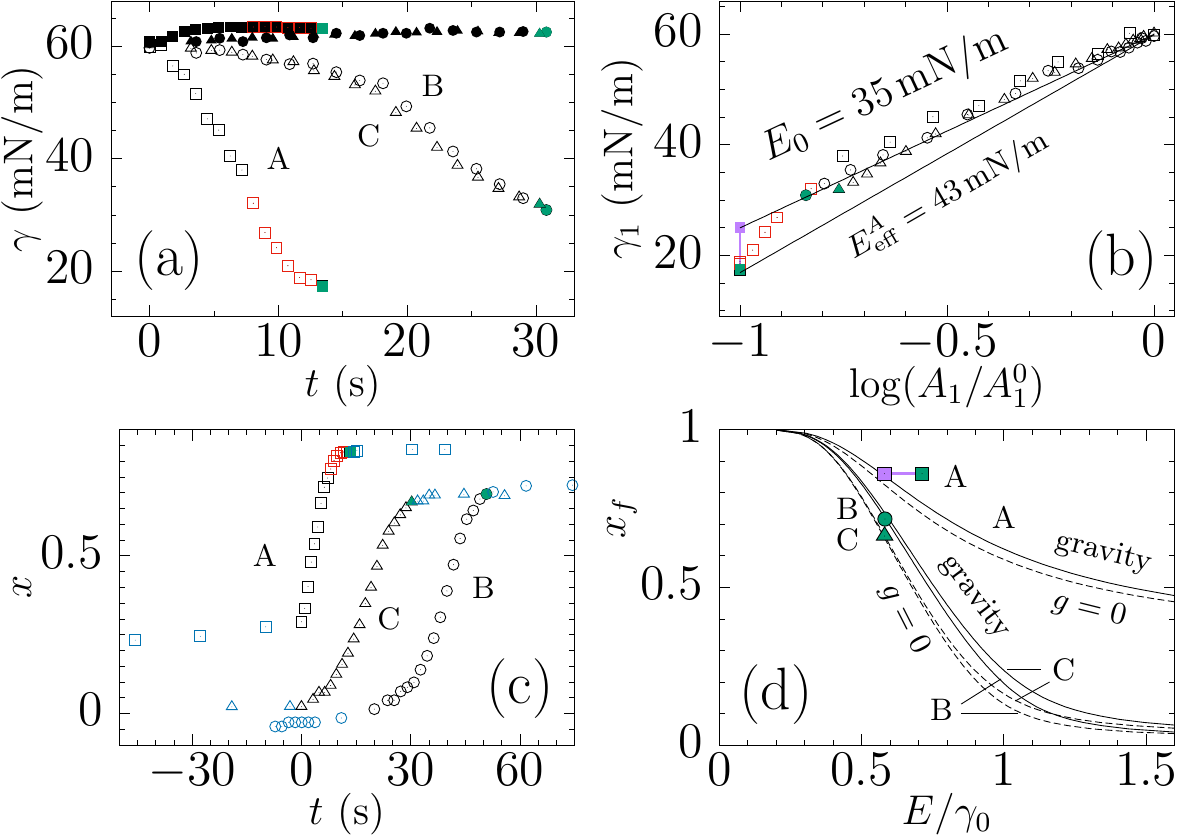}
}
\caption{Evolution of the bubbles in time. The colour and symbol code is the same as in Fig.~\ref{fig:Setup}. {\em (a)} Surface tensions (mN/m) extracted from bubble shapes and sizes. 
{\em (b)} Surface tension $\gamma_1$ of the small bubbles as a function 
of the logarithm of their surface areas $A_1$. 
Two straight lines are plotted as guides for the eye. 
The first one is an estimate of 
the surface elasticity $E_0$ at large $A_1$. 
The second one is the chord of the curve obtained for experiment~A. 
It gives an estimate 
of the effective surface elasticity $E_{\rm{eff}}^{\rm A}$. 
{\em (c)} Evolution of the polydispersity index $x$ defined by Eqs. \ref{eq:R1:x:R0} 
and \ref{eq:R2:x:R0} for all three experiments. 
{\em (d)} Final polydispersity index $x_f$.
The curves represent Eq.~\eqref{eq:xf:predicted:cas:3} 
with the parameters given in Table \ref{Table:Measurements}
while dashed curves are obtained with zero gravity. 
Parameter $x_0$ is measured at the first black point.
The data points correspond to the value of $x_f$
measured at time $t_f$ (green points). 
For experiment~A, 
the purple square corresponds to $E=E_0$
while the green square corresponds to $E=E_{\rm{eff}}^{\rm A}$, as} in \em{(b)}.
\label{fig:courbes:experimentales}
\end{center}
\end{figure}

The evolution of the measured surface tensions are plotted in FIG.~\ref{fig:courbes:experimentales}a. 
At time $t=0$, when the bubbles are put in contact, the surface tensions of both bubbles 
are roughly equal ($\gamma_1=\gamma_2\approx 60$ mN/m for all three  
experiments, see FIG.~\ref{fig:courbes:experimentales}a, in line with \cite{Maestro2014}) 
and stationary, which indicates that, at that time, each bubble has reached equilibrium.

FIG.~\ref{fig:courbes:experimentales}a shows that the surface tension $\gamma_1$ (open, dotted symbols) 
of the smaller bubbles decreases with time  once the valve between the two bubbles has been opened. 
If $E$, defined by equation~\eqref{SurfaceElasticity}, is constant, 
then this equation can be integrated, leading to:
\bee
\gamma_1(t) = \gamma_0 + \elasticity \ln(\area_1(t)/\area_1^0).
\label{eq:SurfaceTension1}
\eee
$\gamma_1$ is thus plotted versus $\ln({A_1}/{A_1^0})$ in FIG.~\ref{fig:courbes:experimentales}b. 
The variation is rather linear, showing that the elasticity $\elasticity$, 
can be considered as both reproducible and almost independent of the surface area. 
It is equal to $\elasticity_{0}=35$~mN/m. 
Note that for experiment A, the slope becomes 
larger ($E\simeq 80$~mN/m) for small surface areas $A_1$
corresponding to tensions below $30$~mN/m.
This probably results from 
the particles starting to be in contact with each other. 
This led us to define an effective elasticity $\elasticity_{\rm eff}^A=43$~mN/m, 
which is the value of $E$ defined by Eq.~\eqref{eq:SurfaceTension1} evaluated at the final time 
considered for experiment A (green point in FIG~\ref{fig:courbes:experimentales}). 
In the following, we will use $E_{0}$ for experiments B and C and both $E_{0}$ and $E_{\rm eff}^{\rm A}$ 
for experiment A.

By contrast, the surface tension of the bigger bubbles is almost constant during the three experiments 
(filled symbols in FIG.~\ref{fig:courbes:experimentales}a), 
which suggests that particle adsorption takes place on much shorter timescales 
than those of the experiment so that the surface elasticity of these bubbles can be neglected. 

In other words, 
the difference between the behaviour of the surface tension 
$\gamma$
of the smaller bubbles
(which shrink)
and the bigger bubbles 
(which expand)
reveal
an assymetry between (slow) desorption 
and (fast) adsorption. 
 
\section{Two bubble coarsening model}

We will now propose a model to describe the above observations.
Building on the experimental observations concerning the asymmetry between adsorption and desorption, 
we propose to write the surface tensions of both bubbles with different expressions. 
The surface tension $\gamma_1$ is given by Eq.~\eqref{eq:SurfaceTension1} 
where $E$ is taken as either $E_0$ or $\elasticity_{\rm eff}^A$. 
The surface tension of the big bubble can be considered as constant:
\be
\gamma_2(t) = \gamma_0.
\label{eq:SurfaceTension2}
\ee

We can now write the pressure in each bubble. The latter must include the effect of gravity.
In the experiment, the syringe outlets are positioned at the same altitude. 
As a consequence, the average altitude of the big bubble
is higher than that of the small one, 
which increases the pressure difference between them and accelerates the coarsening process.
Correspondingly, 
we take into account both the Laplace pressure and the hydrostatic pressure 
in writing the pressure $\pressure_i(t)$ in bubble $i$ as%
\footnote{
Eq.~\eqref{eq:pressure:spherical:laplace:gravity}
for the gas pressure in the bubble was obtained in the following way.
The shape of a bubble of fixed volume $V_i=\frac{4\pi}{3}R_i^3$
attached to an outlet of radius $r_{\rm out}<R_i$
was solved numerically in the presence of gravity $g$.
Then, the first order term in $g$ was retained 
from the limit of zero gravity ($\rho g R_i^2/\gamma_i \rightarrow 0$).
Finally, the outlet radius was taken to the zero limit ($r_{\rm out}\rightarrow 0$).
This procedure sets the coefficients in Eq.~\eqref{eq:pressure:spherical:laplace:gravity}
unambiguously. The same limit was also obtained analytically, 
see Supplementary Material freely available as [????].}:
\bee
\pressure_i(t) &=& 2\gamma_i(t) / R_i(t) -4\rho g R_i(t)/3.
\label{eq:pressure:spherical:laplace:gravity}
\eee
where the reference pressure
is that of the liquid near the needle outlets
and where $R_i$ is the radius of bubble $i$ in the spherical approximation,
defined from its actual volume: $\frac{4\pi}{3}R_i^3=V_i$.

\begin{figure}[h!]
\begin{center}
\resizebox{0.8\columnwidth}{!}{
\includegraphics{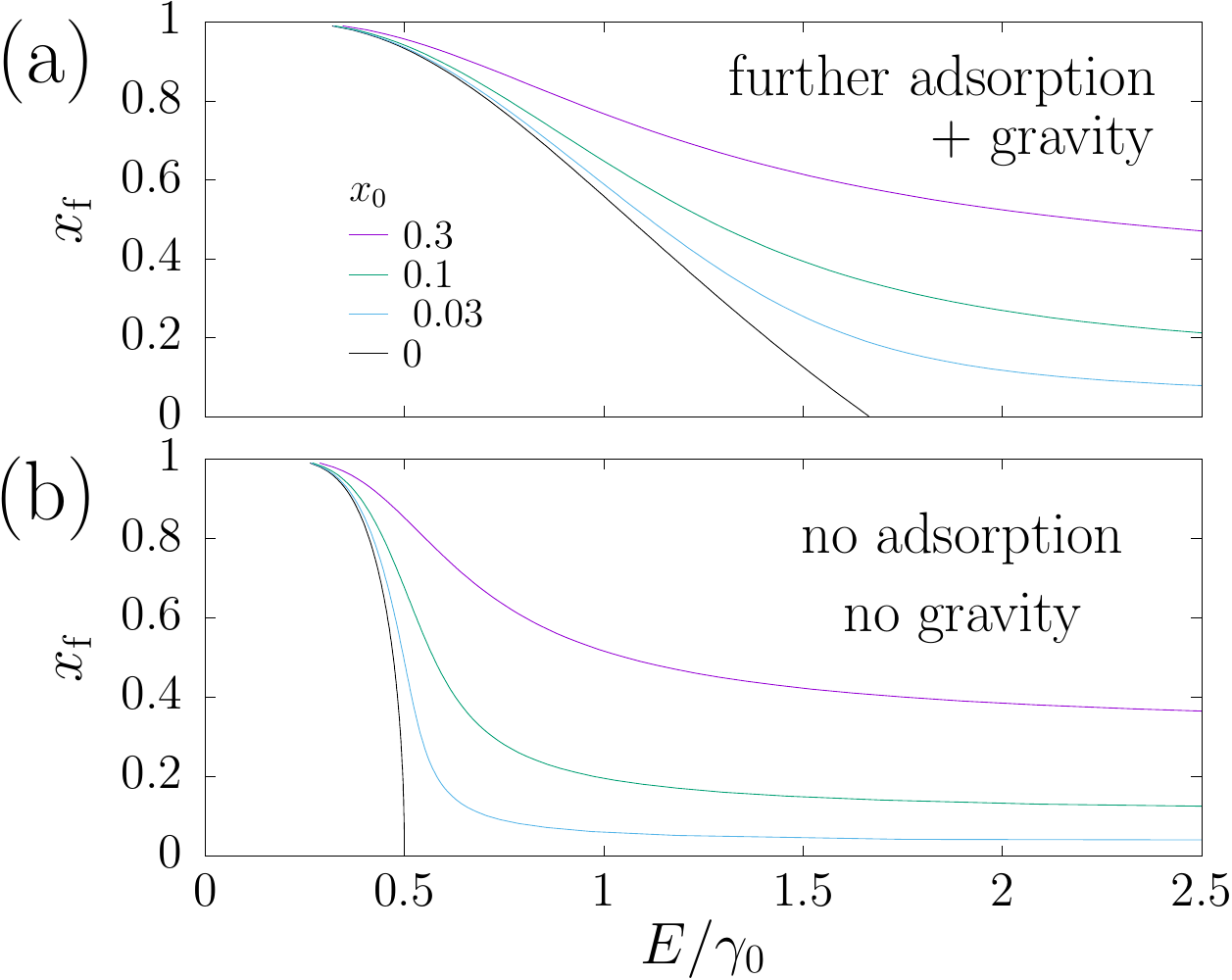}
}
\end{center}
\caption{Final polydispersity $x_f$ of the bubble volumes: 
expected effect of surface elasticity, adsorption and gravity
for irreversibly adsorbed surfactants.
The final volume of the big (resp. small) bubble
is proportional to $1+x_f$ (resp. $1-x_f$).
Black curves correspond to initially identical bubbles
(initial polydispersity factor $x_0=0$)
while coloured curves correspond to $x_0=0.03$, $0.1$ and $0.3$.
(a) Prediction of Eq.~\eqref{eq:xf:predicted:cas:3}.
Further adsorption onto the big (growing) bubble
is assumed to be present.
Gravity is taken into account
and the ratio of the average bubble radius $R_0$
to the capillary length $\lcap=\sqrt{\gamma_0/(\rho g)}$
is taken equal to 1.0.
(b) Prediction of Eq.~\eqref{eq:xf:predicted:cas:1}.
The effect of gravity on the pressure is neglected
and it is assumed that there is no further adsorption onto the big (growing) bubble.
} 
\label{fig:theorie}
\end{figure}

Let us now introduce the effective average radius $R_0$ 
and the polydispersity factor $x$ through
$2R_0^3 = R_1^3+R_2^3$
and $2x = (R_2^3-R_1^3)/R_0^3$, which yields:
\bee
R_1 &=& (1-x)^{1/3}\,R_0
\label{eq:R1:x:R0}
\\
R_2 &=& (1+x)^{1/3}\,R_0,
\label{eq:R2:x:R0}
\eee
where $x=0$ if the bubbles have the exact same size
while $x\rightarrow 1$ in the limit where the smaller bubble shrinks entirely and disappears. 
In the three experiments presented, the initial value $x_0$ of the polydispersity factor 
is varied in the range accessible in the experiment ($x_0=0.01-0.3$, FIG.~\ref{fig:courbes:experimentales}c). 
For simplicity, we assume that all of the gas leaving bubble~1
is transferred to bubble~2 entirely, 
which is reasonable as the volume variation after 
the opening of the valve is much larger than 
that due to the bubble dissolution into the solution (see FIG.~\ref{fig:Setup}). 
Moreover, the gas is considered as incompressible because the pressure difference 
between the two bubbles $p_1-p_2 \simeq 120$ Pa for a bubble radius of $1$ mm 
is much smaller than the pressure itself $p_1 \simeq p_2 \simeq 10^5$ Pa.
As a result of these assumptions, the total volume remains constant, {\em i.e.} $R_0(t)=R_0$.
Using $\area_1=4\pi\,{R_1}^2$ as an approximation of the small bubble surface area,
and Eqs.~\eqref{eq:SurfaceTension1}, \eqref{eq:SurfaceTension2}, 
\eqref{eq:pressure:spherical:laplace:gravity}, 
\eqref{eq:R1:x:R0} and~\eqref{eq:R2:x:R0},
the pressure in each bubble can be written as:
\bee
\pressure_1 &=& \frac{2\gamma_0 
+ \frac43 \elasticity \ln\left(\frac{1-x}{1-x_0}\right)}
{R_0\,(1-x)^{1/3}}-\frac{4}{3}\rho g R_0\,(1-x)^{1/3}.
\label{eq:p1:cas:1}
\\
\pressure_2 &=& \frac{2\gamma_0}{R_0\,(1+x)^{1/3}}-\frac{4}{3}\rho g R_0\,(1+x)^{1/3}
\label{eq:p2:cas:1}
\eee
At equilibrium, both pressures are equal,
hence Eqs.~\eqref{eq:p1:cas:1} and~\eqref{eq:p2:cas:1}
yield a prediction, in implicit form, for the final polydispersity $x_f$
as a function of two control parameters,
namely the initial polydispersity $x_0$ and the surface elasticity $\elasticity$:
\bee
\frac{\elasticity}{\gamma_0}\,
\frac{\ln\left(\frac{1-x_0}{1-x_f}\right)}{(1-x_f)^{\frac13}}
&=& \frac{3}{2} \left[
\frac{1}{(1-x_f)^{\frac13}}
-\frac{1}{(1+x_f)^{\frac13}}
\right]
\fin
&&+\frac{R_0^2}{\lcap^2}
\left[ (1+x_f)^{\frac13} -(1-x_f)^{\frac13} \right].
\quad\quad
\label{eq:xf:predicted:cas:3}
\eee
Here the capillary length is defined as $\lcap=\sqrt{\gamma_0/(\rho g)}$.
The prediction for the final polydispersity $x_f$ is plotted in FIG.~\ref{fig:theorie}a 
for various initial polydispersities $x_0$ for $R_0=\lcap$.

In this case, Eq.~\eqref{eq:xf:predicted:cas:3} predicts 
that if the initial polydispersity $x_0$ is rigorously zero (black line in FIG.~\ref{fig:theorie}a), 
no coarsening occurs ($x_f=0$) whenever $\elasticity/\gamma_0 \geq 5/3$. 
If $\elasticity/\gamma_0 \leq 5/3$, 
the situation is metastable and any perturbation 
will lead to disproportionation. This prediction looks like a modified Gibbs criterion: 
if the bubbles are initially monodisperse, a high enough surface elasticity prevents coarsening. 
We will come back to this observation later on to discuss why we find a criterion of $5/3$ instead of 1/2. 
By contrast, coarsening is expected as soon as $\elasticity/\gamma_0 < 5/3$ 
although it should stop after a finite change in bubble volume, 
i.e. for some final polydispersity $x_f$ strictly between $0$ and $1$. 

If the initial bubbles have slightly different volumes ($x_0>0$), 
the criterion is less sharp (coloured lines in FIG.~\ref{fig:theorie}a): 
the larger the surface elasticity, the smaller the final polydispersity $x_f$. 
Note that this is in qualitative agreement with what we observe in the experiment: 
the coarsening is observed at the beginning but stops definitely 
at a finite polydispersity ({\em i.e.,} the size of the smaller bubble remains finite).

Let us now draw a quantitative comparison between the experiment and the model
given by Eq.~\eqref{eq:xf:predicted:cas:3}.
The initial and final values of the bubble volumes $V_1$ and $V_2$, 
the corresponding initial and final polydispersity factors $x_0$ and $x_f$ are given in Table \ref{Table:Measurements} for all three experiments, 
together with the value of the initial surface tension $\gamma_0$. 
In all experiments, the initial surface tension $\gamma_0$ is close to 60~mN/m 
and the corresponding capillary length is given by $\lcap = 2.45\,{\rm mm}$. 
The value of $R_0/\lcap$, 
where $R_0$ is the effective average radius,
is also available in Table \ref{Table:Measurements}.
The surface elasticity values $\elasticity = 35\,{\rm mN/m}$ 
and $\elasticity_{\rm{eff}}=43\,{\rm mN/m}$ 
(see FIG.~\ref{fig:courbes:experimentales}b)
yield the respective values $0.58$ and $0.71$ of the ratio $\elasticity/\gamma_0$.
%
\begin{table}
\begin{center}
\small
\begin{tabular}{c|c c|c|c|c|c c|c}
  Exp
  & 
  $V_1^0$
  & 
  $V_2^0$
  & $x_0$
  & 
  $\gamma_0$
  & $\frac{R_0}{\lcap}$
  & 
  $V_1^f$
  & 
  $V_2^f$
  & $x_f$ \\
\hline
A & 2.8 & 5.1 & 0.29 & 60.35 & 0.40 & 1.0 & 6.9 & 0.75 \\
B & 3.5 & 3.6 & 0.014 & 60.15 & 0.38 & 1.1 & 6.2 & 0.72 \\
C & 4.5 & 4.7 & 0.022 & 60.3 & 0.42 & 1.5 & 7.6 & 0.66 \\
\end{tabular}
\normalsize
\end{center}
\caption{Different parameters measured for experiements A, B and C: 
initial volumes of the small and big bubbles ($V_1$ and $V_2$ respectively, expressed in mm$^3$), 
corresponding initial and final polydispersities ($x_0$ and $x_f$), 
initial surface tension $\gamma_0$ (mN/m),
average effective bubble radius $R_0$ made dimensionless 
by the capillary length $\lcap = 2.45\,{\rm mm}$.}
\label{Table:Measurements}
\end{table}
In Figure \ref{fig:courbes:experimentales}d, 
the values of $x_f$ extracted from each experiment (see Table \ref{Table:Measurements})
are compared to the theory and agree remarkably well. 
For experiment~A, theory and experiments are compatible
within the error bar due to $\elasticity$.
Note that with the present situation $R_0/\lcap\approx 0.40$,
gravity has very little influence on $x_f$.\newline

\section{Back to Gibbs criterion}

Let us now go back to the (two-bubble) criterion we found and 
see how it compares to Gibbs (one bubble) prediction.
Our calculation predicts no coarsening if 
{\em (i)} the bubbles are initially totally monodisperse 
and {\em (ii)} $\elasticity/\gamma_0 \geq 5/3$. 
This criterion is 
similar but not identical to Gibbs criterion, 
which predicts no coarsening if $\elasticity/\gamma_0 \geq 1/2$. 
This is because we took into account gravity and we supposed a fast adsorption during the growth of the big bubble. 

If we now neglect the effect of gravity and adsorption 
(implying that both bubbles have the same elasticity), 
the hydrostatic pressure has to be removed from Eq.~\eqref{eq:pressure:spherical:laplace:gravity}, and Eq.~\eqref{eq:SurfaceTension2} needs to be replaced by 
\bee
\gamma_2(t) = \gamma_0 + \elasticity \ln(\area_2(t)/\area_2^0).
\label{eq:SurfaceTension2Bis}
\eee
where $\area_2^0$ is the initial area of the bubble $2$ 
and $\area_{2}(t)$ its evolution with time. 

Under such circumstances, 
Eq.~\eqref{eq:xf:predicted:cas:3} is modified 
($\ell_{\textrm{cap}} \rightarrow \infty$)
and the condition $\pressure_1=\pressure_2$ now reads:
\bee
\frac{\elasticity}{\gamma_0}\,
\left[
\frac{\ln\left(\frac{1-x_0}{1-x_f}\right)}{(1-x_f)^{1/3}}
-\frac{\ln\left(\frac{1+x_0}{1+x_f}\right)}{(1+x_f)^{1/3}}
\right]\quad\quad &&
\fin
= \frac32 \left[
\frac{1}{(1-x_f)^{1/3}}
-\frac{1}{(1+x_f)^{1/3}}
\right]. &&
\label{eq:xf:predicted:cas:1}
\eee
This modifies the prediction for the final volume of the two bubbles 
(as shown in FIG.~\ref{fig:theorie}b). 
The result is qualitatively the same, however
{\em (i)} the Gibbs criterion becomes $\elasticity / \gamma_0 \geq 1/2$ for $x_0=0$
and {\em (ii)} for $\elasticity / \gamma_0 <1/2$, 
the final polydispersity factor $x_f$ 
increases more sharply with a decreasing surface elasticity. 
Thus, we recover Gibbs prediction in the absence of gravity and with initially monodisperse bubbles 
provided the surface elasticity is identitical for shrinking or growing bubbles.

More generally, taking Eq.~\eqref{eq:xf:predicted:cas:3} with $x_0=0$
in the limit $x_f\rightarrow 0$,
we obtain a new version of the Gibbs criterion:
\be
\frac{\elasticity}{\gamma_0}
\geq k_{\rm ads}
\left( 1+\frac23\frac{R_0^2}{\lcap^2} \right).
\label{NewCriterion}
\ee
where $k_{\rm ads}=1/2$ when no adsorption takes place, as in Eq.~\eqref{eq:xf:predicted:cas:1},
and $k_{\rm ads}=1$ for fast adsorption as in Eq.~(\ref{eq:xf:predicted:cas:3}), 
and where $\ell_{cap} \rightarrow \infty$ in the zero gravity limit.
   
\section{Conclusion}
 
This comparison between a model and a simple experiment 
performed on two interconnected bubbles 
allows to rationalise why the Gibbs criterion describes foams qualitatively well
even if the threshold of $\elasticity/\gamma_0 \geq 1/2$ is not always recovered experimentally. 
We indeed show in this Letter that the Gibbs criterion 
describes well the case of two bubbles stabilised by agents 
which are irreversibly adsorbed to the interface from the beginning, 
in the absence of gravity {\em i.e.}, for small bubbles (where $R_0 \ll \lcap$),
and for initially monodisperse bubbles. 
In this case, the coarsening does not even start if $\elasticity/\gamma_0 \geq 1/2$. 
If the bubbles are initially polydisperse, the criterion is somewhat relaxed: 
the coarsening starts, but it stops before the small bubble can disappear. 
In the presence of gravity or in the presence of fast adsorption, 
the classical threshold value $1/2$ for the Gibbs criterion increases 
and the total arrest of the coarsening is predicted by a new version of Gibbs criterion, given by Eq.~\eqref{NewCriterion}.
We have shown that when all these different effects are taken into account, 
the experiments are well captured by the theoretical description. 
Coarsening indeed stops experimentally after a finite time 
and bubbles reach a finite volume. 
 
This result is different from the one obtained by Taccoen \textit{et al}, 
which is that their results are in contradiction with an elastic description of the bubble shape. 
In our experiment, the coarsening is actually stopped before the crumpling of the bubbles, 
which may explain this discrepancy.

In order to generalise the present study and describe foam coarsening completely, 
additional steps still need to be taken 
because it may differ from the present two-bubble situation
for at least two reasons.
{\em (i)} When a given bubble swells or shrinks, its various facets 
may expand or shrink, depending on the dynamics of the neighbouring bubbles.
This leads to different local surface tensions on a given bubble.
{\em{(ii)}} In the two bubble experiment, the contribution of Gibbs elasticity and film permeability 
are decorrelated, which allows us to investigate the dominant mechanism in the arrest of foam coarsening. 
The permeability should be included in a full foam coarsening model.
More generally, the fact that the coarsening behaviour of two bubbles 
is very different from that of a single bubble, as we have shown,
suggests that coarsening in 3D foams may display complex collective behaviours. 
 
We believe that the physical understanding we have gained from the two-bubble system 
can be translated directly to other systems which undergo grain growth. 
Most realistic systems are likely to have a more complex elastic behaviour than the one considered here. 
Already in the case of bubbles one may think about more complex scenarios, 
for example when considering soluble surfactants with slow desorption/adsorption. 
In this case, the coarsening would not be stopped, 
but the overall coarsening dynamics would be affected. 

In more general terms, our work on bubbles may provide insight and inspiration to advance 
our understanding and control of other systems (emulsions, alloys, ...) which undergo grain growth. 
It puts in evidence the importance of a finite interfacial elasticity, 
which, in different systems may be created by different means. 
For example, in the case of emulsions, this is achieved by the addition of insoluble, 
interfacially active species (particles, polymers or proteins), 
while in the case of alloys a similar effect is obtained by the addition 
of appropriate dopant atoms \cite{Kirchheim2002}. 
It also points towards a range of subtleties which have to be taken into account, 
including the initial polydispersity, (a)symmetry in the elastic response or additional effects such as gravity.

\section*{Acknowledgments}
We would like to acknowledge fruitful discussions with I. Cantat and D. Langevin. 
A. Maestro is grateful to the CNRS for financing his position. 
We acknowledge funding from the European Research Council (ERC Starting Grant 307280-POMCAPS).
\bibliographystyle{unsrt}

\end{document}